\documentclass{ws-ijmpa}

\begin{document}

\markboth{F. Cianfrani, G. Montani}
{Elementary particle interaction from a Kaluza-Klein scheme}

\catchline{}{}{}{}{}
%
%%%%%%%%%%%%%%%%%%%%%%%%%%%%%%%%%%%%%%%%%%%%%%%%%%%%%%%%%%%%%%%%%%%%

\title{Elementary particle interaction from a Kaluza-Klein scheme
}

\author{FRANCESCO CIANFRANI}

\address{ICRA---International Center for Relativistic Astrophysics\\ 
Dipartimento di Fisica (G9),
Universit\`a  di Roma, ``Sapienza'',\\
Piazzale Aldo Moro 5, 00185 Rome, Italy.\\ 
francesco.cianfrani@icra.it}

\author{GIOVANNI MONTANI}

\address{ICRA---International Center for Relativistic Astrophysics\\ 
Dipartimento di Fisica (G9),
Universit\`a  di Roma, ``Sapienza'',\\ 
Piazzale Aldo Moro 5, 00185 Rome, Italy.\\ 
ENEA C.R. Frascati (Dipartimento F.P.N.),
Via Enrico Fermi 45, 00044 Frascati, Rome, Italy.\\
ICRANet C. C. Pescara, Piazzale della Repubblica, 10, 65100 Pescara, Italy.\\
montani@icra.it}

\maketitle

\begin{history}
\received{Day Month Year}
\revised{Day Month Year}
\end{history}

\begin{abstract}

We discuss properties of particles and fields in a multi-dimensional space-time, where the geometrization of gauge interactions can be performed. As far as spinors are concerned, we outline how the gauge coupling can be recognized by a proper dependence on extra-coordinates and by the dimensional reduction procedure. Finally applications to the Electro-Weak model are presented.


\end{abstract}

\ccode{PACS number: 04.50.Cd}

\section{Introduction}
It is well-expected that unifying gravity with other interactions requires a decisive change of perspective on some of our ideas. We can mention String Theory as the most suggestive approach in this direction. 

Here we are going to focus our attention on the Kaluza-Klein (KK) approach \cite{K,Kl} (see \cite{CMM05} for a recent review), since it realizes the issue of the unification, without avoiding one of the most impressive results of Theoretical Physics: the geometrization of the gravitational field. Furthermore,  such models provide a description of gauge bosons as components of the metric tensor, in a space-time with more than 4 dimensions. The presence of such extra-coordinates does not conflict with a 4-phenomenology, as soon as they are compactified to very small distances. This scenario is not so unusual, because we know that in the very Early Universe the compactification of some dimensions is expected. 
Being undetectable in a direct way, we are looking for a phenomenology of these extra-dimensions via the coupling with matter fields.

In particular, for first we are going to present the main features of the KK space-time and on the way gauge bosons arise from off-diagonal metric components. Then, the interaction with classical test bodies will be investigated, by performing the dimensional reductions of equations describing structure-less and rotating test particles in a 5-dimensional KK background. Finally, the interaction between geometry and spinors will be analyzed, stressing how by a phenomenological point of view some standard issue of KK theories can be avoided, as the emergence of huge mass terms. The application to the electro-weak model will be presented, in order to outline and to support this perspective.   

\section{Kaluza-Klein Theories}

The space-time manifold proper of the KK approach is $V^{4}\otimes B^{K}$, $V^4$ being the 4-dimensional space-time, while the additional space $B^{K}$ is
\begin{itemize}
{\item compact, in order to explain its undetectability at energy scales reachable in current experiments
%\begin{equation}\Delta x=\frac{\hbar c}{E}\end{equation}
}
{\item homogeneous \cite{LL}, {\it i.e.}

\begin{equation}
\label{a1} \xi^{n}_{\bar{N}}\partial_n
\xi_{\bar{M}}^{m}-\xi^{n}_{\bar{M}}\partial_n
\xi_{\bar{N}}^{m}=C^{\bar{P}}_{\bar{N}\bar{M}}\xi^{m}_{\bar{P}},
\end{equation}
this way the algebra of Killing vectors $\xi^{m}_{\bar{P}}$ reproduces the one of gauge group having $C^{\bar{P}}_{\bar{N}\bar{M}}$ as structure constants.}
\end{itemize}

Gauge bosons $A^{\bar{M}}_{\mu}$ are inferred from the metric tensor, once its components are arranged as follows

\begin{equation}
\label{c1}
j_{AB}=\left(\begin{array}{c|c}g_{\mu\nu}-\phi^2\gamma_{mn}\xi^{m}_{\bar{M}}\xi^{n}_{\bar{N}}A^{\bar{M}}_{\mu}A^{\bar{N}}_{\nu}
& -\phi^2\gamma_{mn}\xi^{m}_{\bar{M}}A^{\bar{M}}_{\mu} \\\\
\hline\\
-\phi^2\gamma_{mn}\xi^{n}_{\bar{N}}A^{\bar{N}}_{\nu}
& -\phi^2\gamma_{mn}\end{array}\right).
\end{equation}

At the same time, $g_{\mu\nu}=g_{\mu\nu}(x^{\rho})$ can be identified with the four-dimensional metric, while it is obvious that $\gamma_{mn}=\gamma_{mn}(y^r)$ is the one of $B^K$. In general, there is also a conformal factor $\phi=\phi(x^{\rho})$, which behaves as a scalar field after the dimensional reduction. Despite its role can be relevant for some issues of the KK approach, like the stabilization of the extra-dimensions, and for phenomenological implications (for instance it can provide space-time violations of gauge coupling constant \cite{CMM05} or of the Equivalence Principle \cite{LM}), we will set $\phi=1$ in the following.

We reproduce correct transformation properties for such fields when we restrict the general covariance to the invariance under transformations below

\begin{equation}\label{b1}\left\{\begin{array}{c} x'^{\mu}=x'^{\mu}(x^{\nu})\\y'^{m}=y^{m}+\omega^{\bar{N}}(x^{\nu})\xi^{m}_{\bar{N}}(y^{n})
\end{array}\right..
\end{equation}

In order to recover a 4-dimensional variational principle from the Einstein-Hilbert action in 4+K dimensions, $j_{AB}$ is rewritten in terms of 4-objects, like in the expression (\ref{c1}), and the integration on $B^K$ is performed (dimensional reduction procedure). The main achievement of this scenario is that, as soon as the following conditions stand 

\begin{equation}
G=\frac{{}^{(n)}\!G}{V^{K}}\qquad
\int_{B^{K}}\sqrt{-\gamma}[\gamma_{rs}\xi^{r}_{\bar{M}}\xi^{s}_{\bar{N}}]d^{k}y=-V^{K}\delta_{\bar{M}\bar{N}}
\end{equation}

the Einstein-Yang-Mills Lagrangian comes out, {\it i.e.}

\begin{equation}
S=-\frac{c^{3}}{16\pi G}\int_{V^{4}}
\sqrt{-g}\bigg[R+{}^{(K)}\!R'+\frac{1}{4} F^{\bar{M}}_{\mu\nu}F^{\bar{M}\mu\nu}\bigg]d^{4}x,\label{azspl}
\end{equation}

$F^{\bar{M}}_{\mu\nu}$ being the field strength ($F^{\bar{M}}_{\mu\nu}=\partial_\mu A_\nu^{\bar{M}}-\partial_\nu A_\mu^{\bar{M}}+C^{\bar{M}}_{\bar{N}\bar{P}}A^{\bar{N}}_\mu A^{\bar{P}}_\nu$) of gauge fields $A_\nu^{\bar{M}}$. 
${}^{(K)}\!R'$ is the curvature of $B^K$ and the corresponding term in (\ref{azspl}) acts like a cosmological constant in a 4-dimensional point of view. 

This result outlines how the geometrization of Yang-Mills gauge bosons can be obtained within the KK paradigm, as soon as it can be found a compact and homogeneous space $B^K$ reproducing the gauge algebra.

\section{Matter in a KK framework}

The introduction of matter has to be consistent with the geometrization scheme presented. This statement means that by adding free matter fields, their interaction with the geometry must reproduce the well-known coupling with gauge boson fields. Otherwise, some corrections would come out and an explanation for their undetectability should be given.
 
Let us consider the 5-dimensional case, which allows to geometrize a $U(1)$ gauge model, hence electrodynamics. In this scenario, a 5-test particle follows a geodesics trajectory, but we expect it to be endowed with gravitational and electro-magnetic properties, in a 4-perspective. In particular, the latter are induced by its fifth-coordinate dependence. This fact can be seen starting from the geodesics equation, which reads as 

\begin{eqnarray*}
 {}^{(5)}\!u^A\nabla_A {}^{(5)}\!u_B=0\label{5test}
\end{eqnarray*}

${}^{(5)}\!u^A$ being the 5-velocity $\frac{dx^{A}}{{}^{(5)}\!ds}$. When one introduces the 4-velocity and the quantity $q$, related to the fifth component of ${}^{(5)}\!u_A$ by the relation  

\begin{equation}
q=\frac{\sqrt{4G}}{c}mu_5,
\end{equation}

the five conditions (\ref{5test}) can be rewritten, under KK hypothesis, as 
\begin{equation}
 \left\{\begin{array}{c}q=const. \\ u^\nu\nabla_\nu u_\mu=\frac{q}{mc^2}F^{\phantom1\nu}_\mu u_\nu \end{array}\right..
\end{equation}
Hence the dynamics of a charged test particle is inferred by the dimensional reduction of a test one in a 5-dimensional KK background. It is worth noting the link between the charge and the fifth component of the velocity.
Therefore, the introduction of test particles is consistent with the KK geometrization procedure.

Let us now consider the next order of a multi-pole expansion, the pole-dipole case. The dynamics is described by the set of Papapetrou equations \cite{Pap}, {\it i.e.}
\begin{equation*}
\left\{\begin{array}{c}\frac{D}{{}^{(5)}\!Ds}{}^{(5)}\!P^{A}=\frac{1}{2}{}^{(5)}\!R_{BCD}^{\phantom1\phantom2\phantom3 A} \Sigma^{BC}{}^{(5)}\!u^{D}\quad\\\\
\frac{D}{{}^{(5)}\!Ds}\Sigma^{AB}={}^{(5)}\!P^{A}{}^{(5)}\!u^{B}-{}^{(5)}\!P^{B}{}^{(5)}\!u^{A}\\\\
{}^{(5)}\!P^{A}={}^{(5)}\!m{}^{(5)}\!u^{A}-\frac{D\Sigma^{AB}}{{}^{(5)}\!Ds}{}^{(5)}\!u_{B}\quad\\\\
\Sigma^{AB}{}^{(5)}\!u_{A}=0\qquad\qquad\qquad\qquad\end{array}\right.
\label{pap5}
\end{equation*}

where Pirani consistency conditions \cite{P56} have been adopted. Among new quantities proper of the 5-dimensional description, we stressed the role of the 5-spin tensor $\Sigma^{BC}$. In particular, we identify the following quantities \cite{CMM06}

\begin{equation*}
  S^{\mu\nu}=\Sigma^{\mu\nu}\qquad S_\mu=\Sigma_{5\mu},
\end{equation*}
whose behavior is that of 4-objects under the restricted set of coordinate transformation (\ref{b1}).
Furthermore, we make the hypothesis that the mass $m$ can be expressed in terms of the 5-one as ${}^{(5)}\!m=\frac{1}{\sqrt{1-u_{5}^2}}m=\alpha m$.  

We can now start with the dimensional reduction of the full system of equations (\ref{pap5}). 

From Pirani conditions, we get
\begin{equation*}
 \alpha\{S^{\nu\mu}u_{\nu}+S^{\mu}u_{5}\}=0.
\end{equation*}

Using the last relation, derivatives of the spin tensor can be evaluated and the following expressions for the generalized momentum arise
\begin{eqnarray} {}^{(5)}\!P^{\mu}=\alpha^2[P^{\mu}+u_{5}\frac{DS^{\mu}}{Ds}-ekF_{\rho\nu}u^{\rho}S^{\nu\mu}u_{5}]=\alpha^2\widetilde{P}^{\mu}\\
{}^{(5)}\!P_{5}=\alpha^2[mu_{5}-u_{\nu}\frac{DS^{\nu}}{Ds}+ekF_{\rho\nu}u^{\rho}S^{\nu\mu}u_{\mu}]=\alpha^2\widetilde{P}_{5}.
\end{eqnarray}

The dynamics of the spin tensor can now be inferred and it results to be given by equations

\begin{equation}
\frac{DS^{\mu\nu}}{Ds}=\alpha^2[\widetilde{P}^{\mu}u^{\nu}-\widetilde{P}^{\nu}u^{\mu}]+\frac{1}{2}ekF^{\mu}_{\phantom1\rho}(u^{\rho}S^{\nu}+S^{\rho\nu}u_{5})-\frac{1}{2}ekF^{\nu}_{\phantom1\rho}(u^{\rho}S^{\mu}+S^{\rho\mu}u_{5}).
\end{equation}

It is worth noting that the equations for derivatives of $S_\mu$ do not add any new dynamical information to conditions above. This point stresses that the additional components of the spin tensor do not have an independent evolutionary character with respect to 4-dimensional ones.

Hence equations for the momentum can be splitted. In the case $A=5$ one finds

\begin{equation*} 
\frac{D}{Ds}(\alpha^{2}\widetilde{P}_{5}+\frac{1}{4}ekF_{\mu\nu}S^{\mu\nu})=\frac{D}{Ds}q=0,
\end{equation*}
which means that a conserved quantity $q$ has been identified. This quantity reduces to the test-particle charge in the limit of vanishing spin. All these features suggest to identify $q$ with charge of the rotating object and this conclusion will be confirmed by the coupling with the electro-magnetic field in equations for the momentum. In fact, for $A=\mu$ the following equations are obtained 
\begin{equation}
\frac{D}{Ds}\hat{P}^{\mu}=\frac{1}{2}R_{\alpha\beta\gamma}^{\phantom1\phantom2\phantom3\mu}S^{\alpha\beta}u^{\gamma}+qF^{\mu}_{\phantom1\nu}u^{\nu}+\frac{1}{2}\nabla^{\mu}F^{\nu\rho}M_{\nu\rho},
\end{equation}

$\hat{P}^{\mu}$ being $\alpha^{2}\widetilde{P}^{\mu}+\frac{1}{2}ekF^{\mu}_{\phantom1\rho}S^{\rho}$, while $M^{\mu\nu}$ is given by $\frac{1}{2}ek(S^{\mu\nu}u_{5}+u^{\mu}S^{\nu}-u^{\nu}S^{\mu})$. This quantity is coupled with $\nabla^{\mu}F^{\nu\rho}$, thus it can be identified with the electro-magnetic moment. From this identification, one concludes that $S^\mu$ describe a non-vanishing electric moment.

Finally, the full set of equations reads as
\begin{equation*}
 \left\{\begin{array}{c}\frac{D}{Ds}\hat{P}^{\mu}=\frac{1}{2}R_{\alpha\beta\gamma}^{\phantom1\phantom2\phantom3\mu}S^{\alpha\beta}u^{\gamma}+qF^{\mu}_{\phantom1\nu}u^{\nu}+\frac{1}{2}\nabla^{\mu}F^{\nu\rho}M_{\nu\rho}.\\
\frac{DS^{\mu\nu}}{Ds}=\hat{P}^{\mu}u^{\nu}-\hat{P}^{\nu}u^{\mu}+F^{\mu}_{\phantom1\rho}M^{\rho\nu}-F^{\nu}_{\phantom1\rho}M^{\rho\mu}\\
\hat{P}^{\mu}=\alpha^{2}P^{\mu}+u_{5}\frac{DS^{\mu}}{Ds}-ekF_{\rho\nu}u^{\rho}S^{\nu\mu}u_{5}+\frac{1}{2}ekF^{\mu}_{\phantom1\rho}S^{\rho}\\
S^{\nu\mu}u_{\nu}+S^{\mu}u_{5}=0\end{array}\right.,
\end{equation*}
which coincide with Dixon-Souriau equations \cite{Dix,Sou}, giving the dynamics of a rotating body in an external electro-magnetic field. Therefore, the geometrization of the electro-magnetic fields does not modify the dynamics of the moving object, up to the dipole order.

All these conclusions cannot be extended to elementary particles, neither when a classical description for their dynamics can be given. This feature is connected with the factor $\alpha$, which fixes the condition $u_5^2<1$, hence
\begin{equation}
\frac{q}{m}<\frac{\sqrt{4G}}{c}. 
\end{equation}

However, it has be shown that this limit does not come out if a better treatment of the dependence on the fifth coordinate is given \cite{LM}. In fact, the analysis above compels a localization of particles on the fifth coordinate, too; but such a localization cannot be achieved in a low energy description.

\section{Fermions in a KK framework}
The definition of spinors in a multi-dimensional setting requires to substitute Dirac spinors with a representation of the multidimensional Lorentz group 
  
\begin{equation}
\psi(x)\rightarrow {}^{(n)}\!\Psi(x;y).
\end{equation}

In view of the geometrization, one would like to identify extra-dimensional symmetries with internal ones.

In this respect, let us consider a KK manifold, with Killing vectors as k-bein vectors (hence the gauge group and the extra-space have the same number of dimensions). Under KK hypothesis the Dirac equation reads as
\begin{equation}
i\gamma^{\mu}(D_\mu-A_\mu^{(m)}e^m_{(m)}\partial_m)\Psi+i\gamma^mD_m\Psi-m\Psi=0,
\end{equation}
and, as soon as $D_m\Psi=0$, the coupling with gauge bosons is reproduced, {\it i.e.} $  i\gamma^{\mu}A_\mu^{(m)}\Gamma_{(m)}\Psi$.
However, a solution of the massless Dirac equation does not exist on a compact manifold with positive curvature \cite{Mec}. Moreover any mass term is of the order of the compactification scale (KK mass terms). 

A solution of the massless Dirac equation can be found by adding external gauge bosons and this way was adopted in literature \cite{BL}. A part from being away from the spirit of KK theories, whose aim is to geometrize the boson component, this procedure failed in reproducing the Standard Model of elementary particles.

\paragraph{Averaged equations} We propose that the dimensional reduction procedure can have a non-trivial effect, such that 4-spinors able to solve the above issues arise. In particular, the extra-space being undetectable, we suggest that 4-fields come out as solutions of equations of motion after an averaging procedure on $B^K$ \cite{CM08} (for a justification of this statement see the work \cite{CMlett}). For instance, one can consider just an integration on $B^K$ and solve the following equation

\begin{equation}
 \int_{B^K}i\gamma^mD_m\Psi \sqrt{\gamma}d^ky=0.
\end{equation}

This way extra-dimensional symmetries are not preserved as a consequence of the average, but KK mass terms can be suppressed. We are going to show how this procedure works for the geometrization of the $SU(2)$ gauge coupling.

\paragraph{Dirac equation on $S^{3}$}

Let us consider a spinor on $V^{4}\otimes S^{3}$. 
The relic un-broken symmetry group in the full tangent space is $ SO(1;3)\otimes SO(3)$ and therefore we can build up an eight-component representation in the following way
  
\begin{equation}
{}^{(7)}\!\Psi_{r}=\chi_{rs}\psi_{s}\qquad r,s=1,2,
\end{equation}

$\chi=\chi(y)$ being a $ SU(2)$ representation, while $\psi_{s}=\psi_s(x)$ are Dirac spinors.

We look for a solution of the averaged Dirac equation on $S^3$, {\it i.e.}
  
\begin{equation}
\int_{S^{3}} d^{3}y\sqrt{\gamma}\gamma^{(m)}(e_{(m)}^{m}\partial_{m}-\frac{i}{2}\sigma_{(m)})\chi=0.\label{dirint}
\end{equation}

In particular, we take into account the following form for $\chi$
  
\begin{equation}
\chi_{rs}=\frac{1}{\sqrt{V}}e^{-\frac{i}{2}\sigma_{(p)rs}\lambda^{(p)}_{(q)}\Theta^{(q)}(y^{m})}\label{sp}
\end{equation}

with $V$ the volume of $S^3$, $\lambda$ a constant matrix satisfying 
  
\begin{equation}
(\lambda^{-1})^{(p)}_{(q)}=\frac{1}{V}\int_{S^{3}}
\sqrt{-\gamma}e^{m}_{(q)}\partial_{m}\Theta^{(p)}d^{3}y.
\end{equation}

For functions $\Theta$ we assume 
  
\begin{equation}
\label{theta}\Theta^{(p)}=\frac{1}{\beta}c^{(p)}e^{-\beta\eta}\qquad\eta>0,
\end{equation}

$\beta$ being an order parameter. By performing an expansion in $\beta^{-1}$, we find
  
\begin{equation}
\int_{S^{3}} d^{3}y\sqrt{\gamma}e_{(m)}^{m}\partial_{m}\chi=\frac{i}{2}\sigma_{(m)}\chi+O(\beta^{-1})
\end{equation}

thus, the bigger $\beta$ is, the better the spinor approximates the solution of the massless Dirac equation.
 
This way we can control by an order parameter additional terms arising in the extra-dimensional Dirac equation. These terms will provide violations of gauge symmetries, whose conservation can account for its small value.
 
The application of this results to a KK space-time $V^4\otimes S^3\otimes S^1$ allows us to reproduce any lepton family and quark generation from the 2 following multi-dimensional spinors \cite{CM08}

\begin{equation} 
\Psi_{L}=\frac{1}{\sqrt{V\alpha'}}\left(\begin{array}{c} \chi\left(\begin{array}{c} e^{in_{uL}\theta}u_{L}\\e^{in_{dL}\theta}d_{L}\end{array}\right)\\\chi\left(\begin{array}{c} e^{in_{\nu L}\theta}\nu_{eL}\\e^{in_{eL}\theta}e_{L}\end{array}\right)\end{array}\right),\qquad
\Psi_{R}=\frac{1}{\sqrt{V\alpha'}}\left(\begin{array}{c} \left(\begin{array}{c} e^{in_{uR}\theta}u_{R}\\e^{in_{dR}\theta}d_{R}\end{array}\right)\\\left(\begin{array}{c} e^{in_{\nu R}\theta}\nu_{eR}\\e^{in_{eR}\theta}e_{R}\end{array}\right)\end{array}\right), 
\end{equation}  

$\alpha'$ and $\theta$ being the length and the coordinate on $S^1$, respectively, while the number $n$ is connected with the hypercharge $Y$ of the corresponding particle by the relation $n_i=6Y_i$.
 
This way, an explanation for the equal number of quark generations and fermion families is provided.
So doing, we are assuming there exists a symmetry relating leptons and quarks, which is broken by the compactification.

A scalar field can be introduced, which realizes the spontaneous symmetry breaking mechanism to $U(1)$ and gives masses to particles. This approach accounts for
\begin{itemize} 
{\item the stabilization of the Higgs mass, by a KK mass term;}
{\item massive neutrinos.}
\end{itemize}

Estimates for the parameter $\beta$ come out from gauge-violating term. If the Higgs mechanism is not taken into account, the lower bound comes from current limits on the electric charge-violating decay of a neutron \cite{10}, {\it i.e.}
\begin{equation}
\Gamma(n\rightarrow p+\nu_{e}+\bar{\nu}_{e})/\Gamma_{tot}<8*10^{-27}\Rightarrow\beta>10^{14}.  
\end{equation}
Whether we also consider corrections induced on gauge boson masses, we obtain the following estimate from the upper bound on the photon mass \cite{10}
  
\begin{equation}
m_{\gamma}<6*10^{-17}eV\Rightarrow\beta>10^{28}
\end{equation}

\paragraph{Non-local dependence}
 
A different perspective in view of the geometrization of the gauge coupling for spinors is based on giving an exact solutions of the massless Dirac equation. This form must be non-local and it reads as

\begin{equation}
{}^{(n)}\!\Psi(x,y)=e^{\int_{C(y)} \Gamma_mdy^m}\psi(x),
\end{equation}

$C(y)$ being a path in $B^K$ connecting a fixed point with that of coordinates $(y)$. This approach is again under investigation.

%\begin{thebibliography}{000} %for 3 digits
%\begin{thebibliography}{00}  %for 2 digits

\end{document}